\documentclass{sigchi}

\toappear{}

\usepackage{balance}  % to better equalize the last page
\usepackage{graphics} % for EPS, load graphicx instead
\usepackage{times}    % comment if you want LaTeX's default font
\usepackage{url}      % llt: nicely formatted URLs
\usepackage{subfigure}

\makeatletter
\def\url@leostyle{%
  \@ifundefined{selectfont}{\def\UrlFont{\sf}}{\def\UrlFont{\small\bf\ttfamily}}}
\makeatother
\def\pprw{8.5in}
\def\pprh{11in}

\usepackage[pdftex]{hyperref}
\hypersetup{
pdftitle={Simulation and Virtual Prototyping of Tangible User Interfaces},
pdfauthor={Stefan Diewald, Andreas M\"{o}ller, Luis Roalter, Matthias Kranz},
pdfkeywords={TUI prototyping, middleware, virtual TUI, Gazebo, ROS},
bookmarksnumbered,
pdfstartview={FitH},
colorlinks,
citecolor=black,
filecolor=black,
linkcolor=black,
urlcolor=black,
breaklinks=true,
}

\begin{document}

\title{Simulation and Virtual Prototyping\ \\of Tangible User Interfaces}

\numberofauthors{1}
\author{
  \alignauthor Stefan Diewald\,$^{1}$, Andreas M\"{o}ller\,$^{1}$, Luis Roalter\,$^{1}$, Matthias Kranz\,$^{2}$\\
    \affaddr{$^{1}$\,Technische Universit\"{a}t M\"{u}nchen, Arcisstra{\ss}e 21, 80333 Munich, Germany}\\
    \affaddr{$^{2}$\,Universit\"{a}t Passau, Innstra{\ss}e 43, 94032 Passau, Germany}\\
    \email{stefan.diewald@tum.de, andreas.moeller@tum.de, roalter@tum.de, matthias.kranz@uni-passau.de}\\
}

\maketitle

\begin{abstract}
Prototyping is an important part in research and development of tangible user
interfaces (TUIs). On the way from the idea to a working prototype, new hardware
prototypes usually have to be crafted repeatedly in numerous iterations. This
brings us to think about virtual prototypes that exhibit the same functionality as a real TUI, but reduce the amount of time and resources that have to be spent.

Building upon existing open-source software -- the middleware Robot Operating
System (ROS) and the 3D simulator Gazebo -- we have created a toolkit that
can be used for developing and testing fully functional implementations of a
tangible user interface as a virtual device. The entire interaction between the
TUI and other hardware and software components is controlled by the middleware,
while the human interaction with the TUI can be explored using the 3D simulator
and 3D input/output technologies. We argue that by simulating parts of the
hardware-software co-design process, the overall development effort can be
reduced.
\end{abstract}

\keywords{
	TUI prototyping, middleware, virtual TUI, Gazebo, ROS
}

\category{H.5.2}{Information Interfaces and Presentation}{User
Interfaces}[Prototyping]

\section{Introduction}

The development and evaluation of prototypic tangible user interfaces
(TUIs)~\cite{Ishii1997} consumes a lot of effort and time due to iterative
design and debug processes on some kind of hardware. Starting from I/O
cubes~\cite{Schiettecatte:2008:ADC:1347390.1347394}, to
tabletops~\cite{McAdam2011} and various augmented everyday
objects~\cite{Chung2010, Wakkary2007}, each TUI consists of individual hardware
that has often to be built from scratch. In order to reduce the development time
for initial prototypes, hardware frameworks, such as \emph{Blades \&
Tiles}~\cite{Ullmer:BladesAndTiles}, \emph{Pin \& Play}~\cite{1193980} or
\emph{Gadgeteer}~\cite{Villar2011}, are used by developers. However, still a lot
of work has to be spent on the hardware before a running prototype can actually
be used for evaluating user interaction and HCI-related aspects.

Hence, a prototyping approach allowing the simulation of tangible user
interfaces at an early stage, e.g.~to evaluate novel interaction concepts,
before building any kind of hardware could extremely shorten the overall
development time. This is especially the case for TUIs that are based on novel
hardware that is not yet available or cannot be realized within reasonable
expenditure.

In this paper, we introduce a toolkit for TUI simulation that allows shifting
the early prototyping process into a high-fidelity 3D virtual environment~\cite{TEI14_TUI_Simulation}. That
way, shapes of objects for a new TUI and/or new interaction concepts can be evaluated before
an actual hardware prototype needs to be built. The proposed toolkit is based on
a middleware that can be used for virtual as well as for real TUIs. For that
reason, interactions between real \emph{and} simulated components are possible.
Moreover, the transition from the virtual to the real prototype does not entail significant changes from software side.

The paper is structured as follows: We first give an overview of existing
state-of-the-art prototyping tools for TUIs. Then, the requirements for a TUI
simulation environment are developed and presented. Based on these requirements,
the designed TUI simulation toolkit is introduced. In a comparison between the
development process of a real prototype and a virtual prototype, the working
method with the proposed solution is presented. Subsequently, we portray the
challenges that need to be overcome for being able to assess interactive as
well as tangible aspects of a virtual TUI with the proposed toolkit. The paper
concludes with a summary of achievements and describes directions for future
work.

\begin{figure}[t]
 \centering
 \subfigure[Real TUI with a display on each side. Inside the cube are an
accelerometer and a communication device.]{
    \label{fig:tui:real}
    \includegraphics[height=0.218\textwidth]{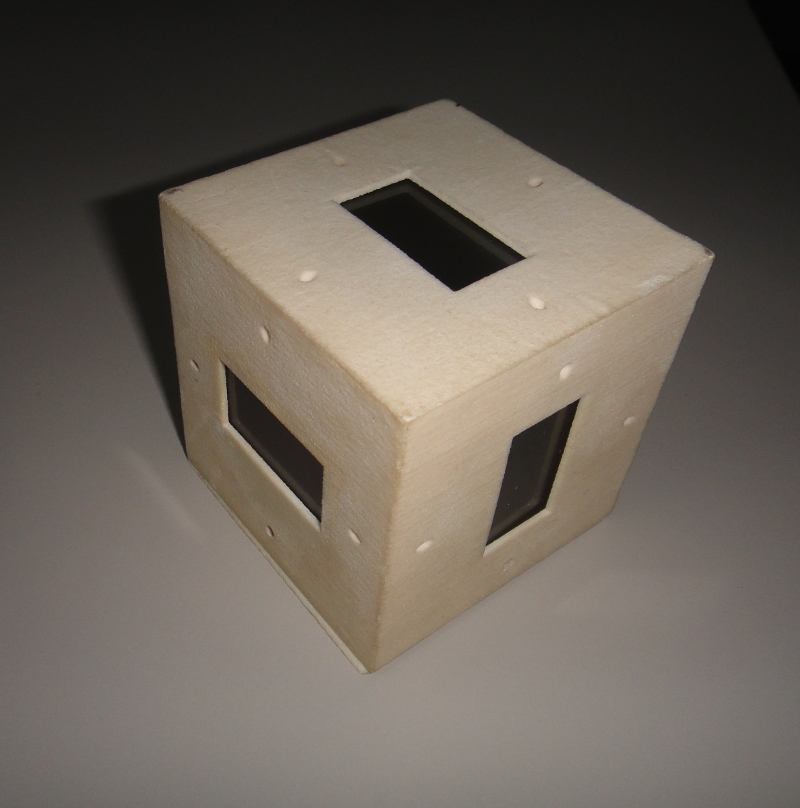} 
  }
  \,
  \subfigure[Virtual TUI simulated with Gazebo in the ROS. The displays are
fully
functional.]{
    \label{fig:tui:virtual}
    \includegraphics[height=0.218\textwidth]{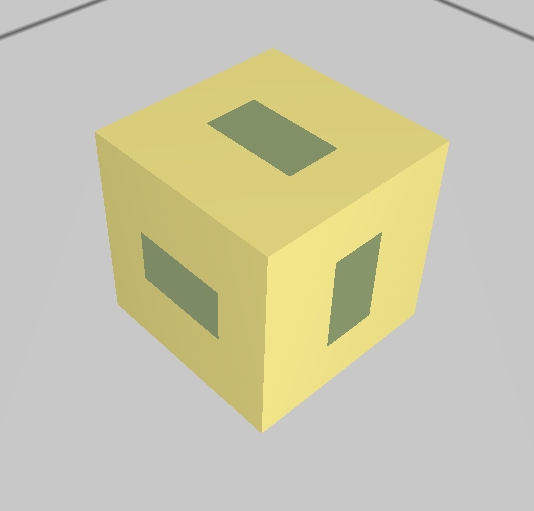} 
  }
  \caption{Both \emph{Display Cube} devices look identically. The only
difference of the cube is in its colors.}
  \label{fig:tui}
\end{figure}

\section{Related Work: Prototyping of Tangible User Interfaces}

Before being able to enhance the process of TUI prototype development and
testing, needs and requirements for TUI development have to be investigated.
We derive these demands by reflecting our own experiences in TUI
prototyping and by examining the features of currently available TUI software
frameworks and hardware toolkits.

\subsection{Tangible User Interface Markup Language (TUIML) and Management
System (TUIMS)}

During the early ideation phase of tangible user interfaces, it is important to
create a clear definition of the interplay between the digital and physical
domain. In 2009, Shaer and Jacob presented the \emph{Tangible User Interface
Markup Language~(TUIML)}~\cite{Shaer2009}. It can be used to describe the
structure and behavior of TUIs.

For creating the high-level description of a TUI in \emph{TUIML}, the \emph{TUI
Management System~(TUIMS) Prism}~\cite{Shaer2009} has been created. It is
an interactive application that allows modeling a TUI and assigning behaviors to
it. A canvas view provides a graphical editor for sketching 3D representations
of the physical objects. The TUI description is automatically translated to
\emph{TUIML}. The additional run-time environment supports a Java3D graphical
simulation of the physical objects. In addition, it can control two
microcontroller platforms and a RFID reader. The source code for the
microcontroller platforms can be generated automatically by the management
system. One goal of such a markup language is to obtain a standardized
description of a TUI which can be used in a collaborative development process.

In contrast to this system, our approach is based on a high-fidelity 3D
simulation connected to a middleware that is equally used for simulated objects
as well as for real physical objects. Our toolkit allows merging the virtual and
the real domain; no distinction is made between real objects and simulated
objects. The real prototype can run the same code, trigger the same actions and
be influenced by the same parameters as the simulated prototype, just by
replacing the device drivers. Our approach does not replace nor contradict the
solution by Shaer et al., but increases the interaction evaluation possibilities
in the early design phase, since the simulation is not any more decoupled from
real objects and real actions.

\subsection{TUI Hardware Toolkits and Software Frameworks}

During the last few years, several hardware toolkits and software frameworks for
TUI prototyping have emerged and therewith simplified the TUI development
process. One of the goals of these frameworks is the decoupling of the hardware
and software development.

The \emph{Blades \& Tiles} from Sankaran et~al.~\cite{Ullmer:BladesAndTiles} is
an example for such a hardware toolkit. By combining different Blades and Tiles
various interaction devices can be realized. \emph{BOXES}~\cite{Hudson2006} is
another rapid construction toolkit for interactive physical prototypes in the
early design stages. With \emph{BOXES}, physical objects can be made interactive
by attaching so-called thumbtacks~\cite{Hudson2006} to them. The little touch
sensors can then be mapped to specific keyboard or mouse input events on the
connected PC.

The \emph{Papier-M\^ach\'e}~\cite{Klemmer2004} toolkit enables to build tangible
interfaces using computer vision, electronic tags, and bar codes. Owners of a
Vicon Tracking system can make use of the \emph{DisplayObjects} rapid
prototyping workbench~\cite{Akaoka2010}. It allows designing functional
interfaces on 3D physical objects of any shape. Other commonly used hardware
toolkits are Arduino\footnote{\url{http://www.arduino.cc/}, last accessed
May\,6, 2014}, Phidgets\footnote{\url{http://www.phidgets.com/}, last
accessed May\,6, 2014} or iStuff~\cite{Ballagas2003}.

\emph{HephaisTK}~\cite{Dumas2009} is an agent-based software toolkit for rapid
creation of multimodal interfaces. The agents are managing the communication
between input recognizers, extraction engines, and output modules. A central
postman is used for storing incoming input events and distributing these
messages to agents that are subscribed to the specific message types.
\emph{Responsive Objects, Surfaces, and Spaces~(ROSS)} API~\cite{Wu2012} and
\emph{reacTIVision}~\cite{Kaltenbrunner2007} are examples for tabletop TUI
prototyping toolkits. \emph{OpenInterface}~\cite{Serrano2008} is a
component-based tool for rapid development of multimodal input interfaces. Its
integrated development environment allows graphical high-level assembling of
different components such as device drivers, interaction techniques, or
multimodal fusion facilities.

\begin{figure}[tb]
  \centering
  \includegraphics[width=0.95\linewidth]{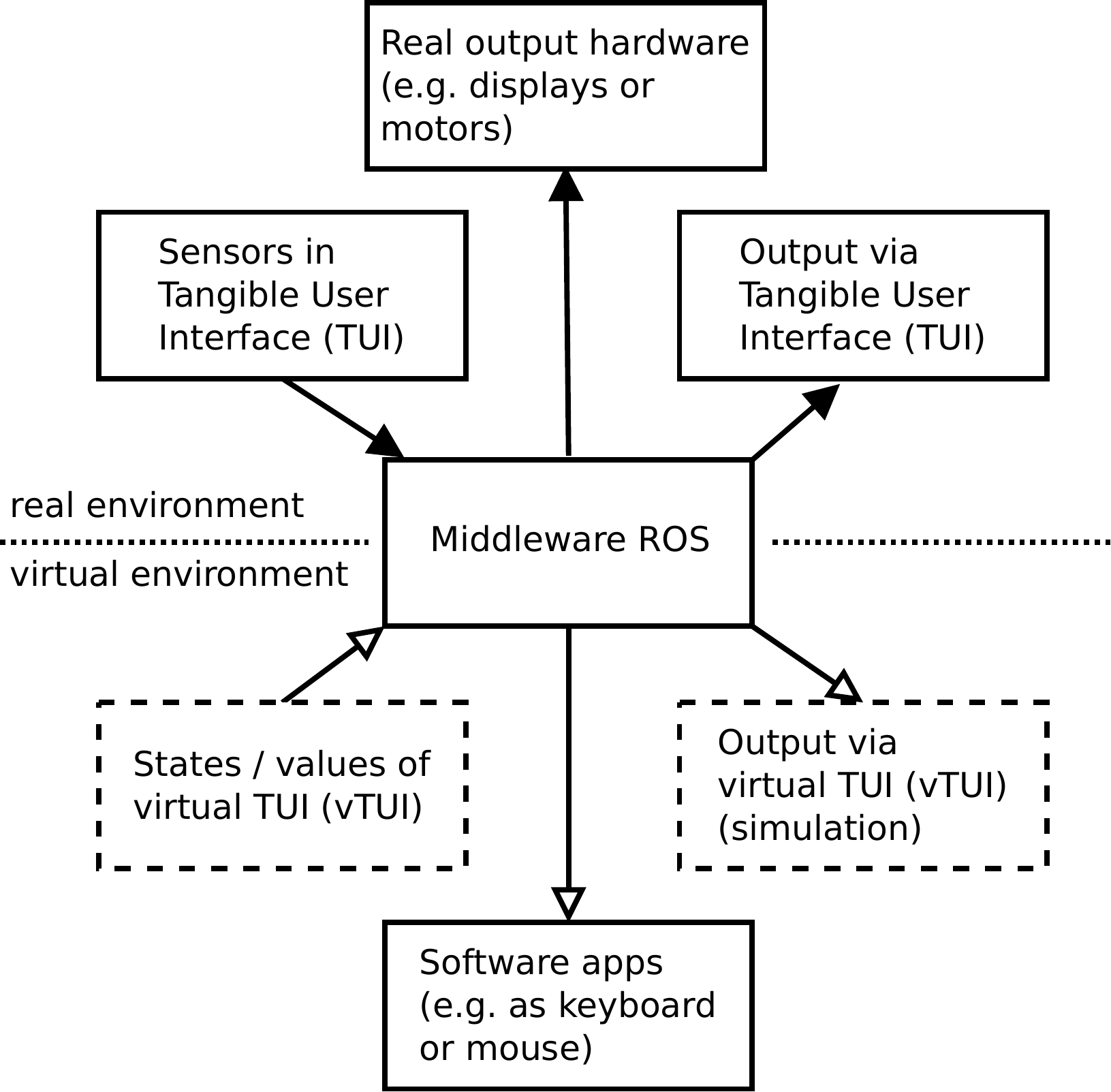}	
  \caption{Up to now, TUIs consisted always of a physical part that only
influenced the virtual environment by controlling software applications running
on a computer. The proposed solution introduces virtual TUIs and proposes a
middleware for connecting elements in real and virtual environments.}
  \label{fig:comparingdevices}
\end{figure}

Low-fidelity prototyping for evaluating tangible interactions is possible with
\emph{Sketch-a-TUI} by Wiethoff et al.~\cite{WiethoffSketchTUIInteraction2012}.
\emph{Sketch-a-TUI} is based on conductive ink that can be applied to cardboard
objects. In this way, the objects can be detected by capacitive touch screens
that, for example, are built-in in state-of-the-art smartphones and tablet PCs.
This allows for fast and cost-efficient prototyping of tangible interactions
since the graspable objects do not need any industrial production process and,
thus, are very cheap. However, the created objects do not have any intelligence
on their own and only work on a very limited space.

\section{Our Approach: Prototyping with Virtual Hardware}

Since all of the currently available TUI prototyping toolkits and methods need
some kind of dedicated real hardware to test the functionality of the TUI, we
have considered a prototyping platform that can be used without any real
hardware components. Based on the results from Kranz et al. with intelligent and smart
environments~\cite{Moller2011,Kranz2011}, we created a toolkit that
allows a complete virtual representation of a TUI. As presented in
Fig.~\ref{fig:tui}, we can create high-fidelity virtual prototypes that look
almost identical to subsequently developed real prototypes.

\subsection{Requirements for a Virtual TUI Simulation Environment}

A simulation environment for TUIs needs to fulfill several requirements in order
to enable a complete evaluation of the system at an early stage and to fulfill
the key properties of TUIs as described by Kim and Maher~\cite{Kim2008222}.
Since tangible user interfaces are based on the linkage of the virtual and physical domain,
it is important that the simulation can simulate any kind of physical object,
especially rigid body objects that are commonly used in many activities of
daily living (ADL). A physics engine has to ensure that the virtual objects
behave like real physical ones. Besides the support for modeling objects of any
arbitrary shape, it should further support realistic textures. Simple geometric
shapes~\cite{KranzDistScroll2005,Wimmer2006Thracker} as well as complex TUIs, such as \emph{topobo}~\cite{Raffle2004}, should
be supported. In order to allow an intuitive evaluation by the user it has to
offer an intuitive 3D interface with the possibility to interact with the
simulated objects and to explore the virtual environment.

Another important factor is the connection of the simulation environment to a
middleware that can actively support the TUI development. It would be useful to
choose one that can be used for the virtual simulation as well as afterwards for
a real implementation.

\subsection{The ROS Middleware as TUI Middleware}

Based on former research on the simulation of intelligent environments, we
have chosen the \emph{Robot Operating System} (ROS) as middleware. ROS is
one of
the major middleware systems in the domain of robotics, e.g.~running on the PR2
and several other robots~\cite{Cousins2010}. Thus, an advantage of this middleware is that a huge set of drivers and applications
(mainly for robotic systems) is already available.
ROS has also been used on immobile robots \emph{(ImmoBots)} such as intelligent environments~\cite{Roalter2011}. 
As argued in this work, intelligent objects behave somewhat like robots as well. This
allows us to deal with TUIs as if they were robots, implying that we are
able to use any robotic simulation method equally for TUIs.

As presented in Fig.~\ref{fig:comparingdevices}, the middleware brings
different hardware and software concepts together, creating the possibility to
interconnect real TUIs with virtual TUIs (vTUIs).

Using the same toolkit for intelligent environments, robots and TUIs, a common
middleware reduces the amount of code that has to be written to establish the
communication between these kinds of systems. The middleware already provides us
with basic messages and communication protocols to transfer any kind of data
between different nodes. For the virtual development process, ROS does not
depend on existing hardware. The developers are completely free in designing the
communication with other TUIs and \emph{Smart Things}.

\begin{figure*}[ht]
  \centering
  \includegraphics[width=0.8\textwidth]{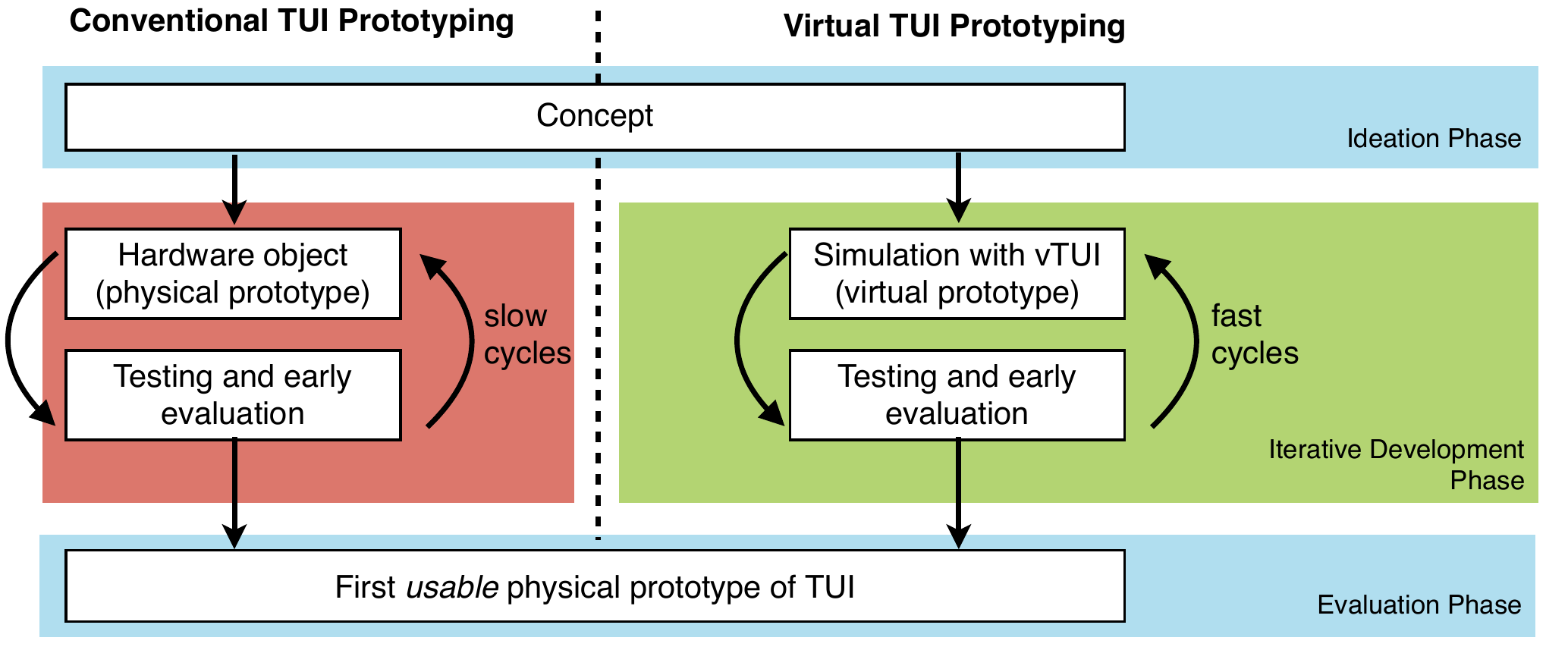}
  \caption{Comparison of the prototyping with real (left)
and virtual (right) devices from the idea to the first physical prototype that
can be used for a user evaluation.
As potentially multiple physical prototypes are needed when prototyping with real hardware, development cycles are faster with virtual prototyping.}
  \label{fig:prototypingphases}
\end{figure*}

\subsection{3D Simulation of Physical Objects with Gazebo}

The 3D simulation is performed with Gazebo~\cite{Koenig2004}, a 3D robot
simulator.
Gazebo is a complete physical simulation of robots including shapes, joints,
contacts, collisions, and friction. Gazebo utilized the 3D framework \emph{OGRE
(Object-Oriented Graphics Rendering
Engine)}\footnote{\url{http://www.ogre3d.org/},
last accessed May\,11, 2014.} for rendering the environment and objects. It
uses the \emph{Open Dynamics
Engine (ODE)}\footnote{\url{http://ode-wiki.org/wiki/}, last accessed May\,11, 2014.} library as physics engine that can simulate rigid body dynamics.
During the simulation, Gazebo publishes the models' states and behaviors via
the ROS middleware's communication infrastructure so that all other nodes can
use these parameters for triggering certain actions.

Gazebo uses URDF (Unified Robot Description
Format)\footnote{\url{http://www.ros.org/wiki/urdf}, last accessed May\,12,
2014.} files for the description of
the models. The physical elements for the simulation can be modeled with all
common 3D modeling tools such as
\emph{Blender}\footnote{\url{http://www.blender.org/}, last accessed May\,12,
2014.},
\emph{Cinema4D}\footnote{\url{
http://www.maxon.net/products/cinema-4d-studio.html } ,
last accessed May\,12, 2014.},
\emph{Autodesk Maya}\footnote{\url{http://usa.autodesk.com/maya/}, last accessed
May\,12, 2014.} or \emph{3ds
Max}\footnote{\url{http://usa.autodesk.com/3ds-max/},
last accessed May\,12, 2014.}. Gazebo
further offers the possibility to draw defined simple geometric objects, such as
boxes, spheres, and cylinders with a single XML tag. All of these simple
elements or meshed elements, i.e.~more complex polyhedral objects, can be linked
together with joints. Gazebo distinguishes between different kinds of joints to
be able to calculate their physical state. This enables to confine the degrees
of freedom to the desired ones. After choosing a type of joint (e.g.~prismatic,
revolute, or sliding), the limits of the joint and the forces needed to
interact with this joint in the simulation can be set. By defining a mass,
inertia and friction values for an element, the physics engine can simulate its
dynamic behavior in a realistic way.

Originally developed as outdoor robotic sensor simulator, a specialty of Gazebo
are virtual sensors and actuators that can be assigned to any object in the
simulation. Examples of available sensors are cameras, laser scanners, contact
switches, force sensors, or inertial measurement units~(IMUs). It is even
possible to simulate a simple battery unit that can be loaded and drained, which
is another important factor for modeling wireless active components for TUIs.
Diewald et al.\,have presented a more extensive list of available sensors and
actuators for Gazebo~\cite{MobileSimulation11}.

Gazebo's functionality can be easily extended through a well-documented API.
For example, we have added the support for touch-sensitive virtual
displays. Based on this extension, we are able to simulate complete tabletop
TUIs or TUIs with embedded displays such as the \emph{Display
Cube}~\cite{kranz2005display} (see Fig.~\ref{fig:tui}) which is used for comparing the development
process of a real TUI to a simulated TUI in a later section of this paper.

For testing and evaluating a TUI system, users can interact with the physical
objects through a GUI. They can apply rotational and translational force to any
object in the simulation.
 
\subsection{Combination of Real and Virtual TUIs}

Hardware abstraction allows using the ROS as middleware for real as well as for
simulated TUIs. A common abstract hardware layer is used for hiding the actual
implementation and for handling the exchange of states and values. The states and
values can be accessed by publishing/subscribing to nodes, or requested as part of a
service callback. By recording and afterwards playing back the messages the exchanged messages,
the middleware can simulate individual objects and larger parts of the
environment or setup, without the need of performing input actions repeatedly. By using
high-level hardware layer bindings, the injection of hardware messages is
possible with little effort. Connecting virtual and real TUIs to the same
middleware network transparently joins both virtual and real hardware together.
They are indistinguishable for other nodes.

\section{Comparison of the Development Process of a Real and a Simulated
Tangible User Interface}

Most of the differences between performing rapid prototyping on real hardware
and modeling a system virtually emerge during the early prototyping phase. We
illustrate the advantage of virtual modeling over conventional modeling by
comparing the development process of a cube with a 3D accelerometer and six
displays as an example for a TUI.
For the real prototype, one needs to choose the
sensors and hardware components which fulfill the needs of the developer. This
process is time-consuming. Using a virtual TUI, the developer only needs to
specify the parameters s/he needs to get from the TUI, such as pose, location, etc.
In the simulation, the desired actions can be attached to these parameters, so that 
the simulator can be used for evaluating the model and the designed behavior.
This approach helps finding the necessary and proper
parameters before the real prototype is implemented.
A last goal of such an approach is to separate physical development from the software development.

\begin{figure*}[htb]
  \centering
  \includegraphics[width=0.98\textwidth]{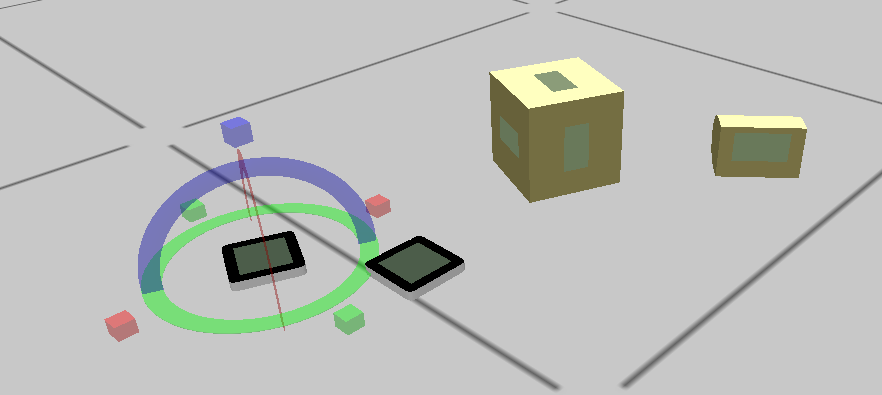}	
  \caption{Virtual TUIs. From left to right: two \emph{Sifteo cubes}, an
I/O cube and a ubiquitous presence system. The left vTUI is currently
manipulated through the GUI. Gazebo allows for manipulation in all 6 degrees of
freedom by applying force to the object.}
  \label{fig:virtual_tuis}
\end{figure*}

A typical iterative development model for the prototyping process is shown in
Fig.~\ref{fig:prototypingphases}. It is barely possible to meet all
requirements for the actual implementation with the first prototype. Hence, the
cycle needs to be traversed multiple times to iteratively improve the prototype.
Creating and refining virtual models is usually significantly faster than creating physical prototypes,
which allows for faster iterations in the development cycle of prototyping and testing.
Often, detected deficiencies after testing result in the creation of an entirely new model. In the virtual
representation, the object of the last iteration can much easier be reused by
e.g.~modifying the shape or texture.
Using the ROS for connecting the devices, one can simultaneously develop the software for the virtual device and for the
final prototype.

In currently available toolkits and TUIs, a communication protocol to connect
heterogeneous hardware has to be created explicitly. The ROS middleware
simplifies this process by reducing the code that has to be written just to a
device driver that connects the device to an actual ROS node. With increasing
processing power of microcontroller platforms, the ROS node could in the future
reside in the TUI itself.

\begin{figure}[tb]
  \centering
  \includegraphics[width=0.5\textwidth]{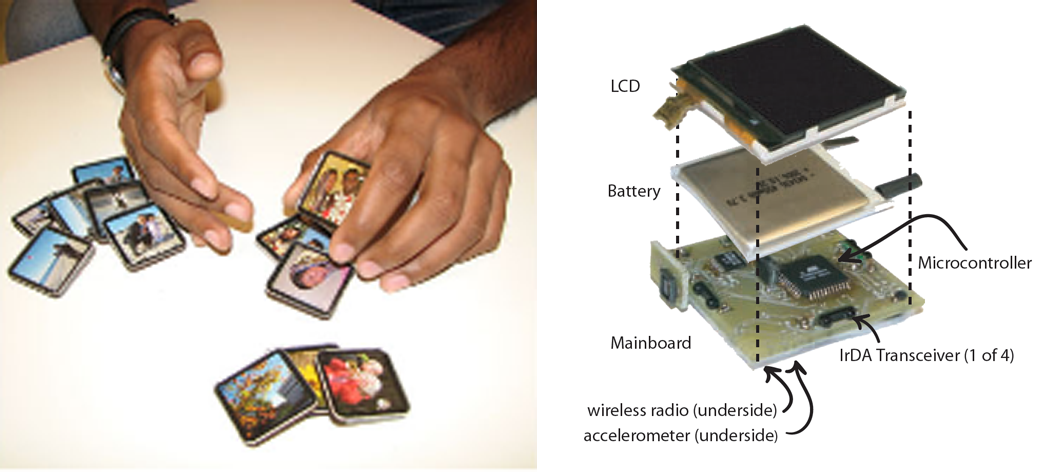}	
  \caption{Left: The real counterpart to the virtual \emph{Sifteo cubes}
depicted in Fig.~\ref{fig:virtual_tuis}. Right: A view on the hardware of the
cubes. \mbox{Image source: Merrill et al.~\protect\cite{Merrill2007}.}}
  \label{fig:real_sifteo}
\end{figure}

\begin{figure}[tb]
  \centering
  \includegraphics[width=0.48\textwidth]{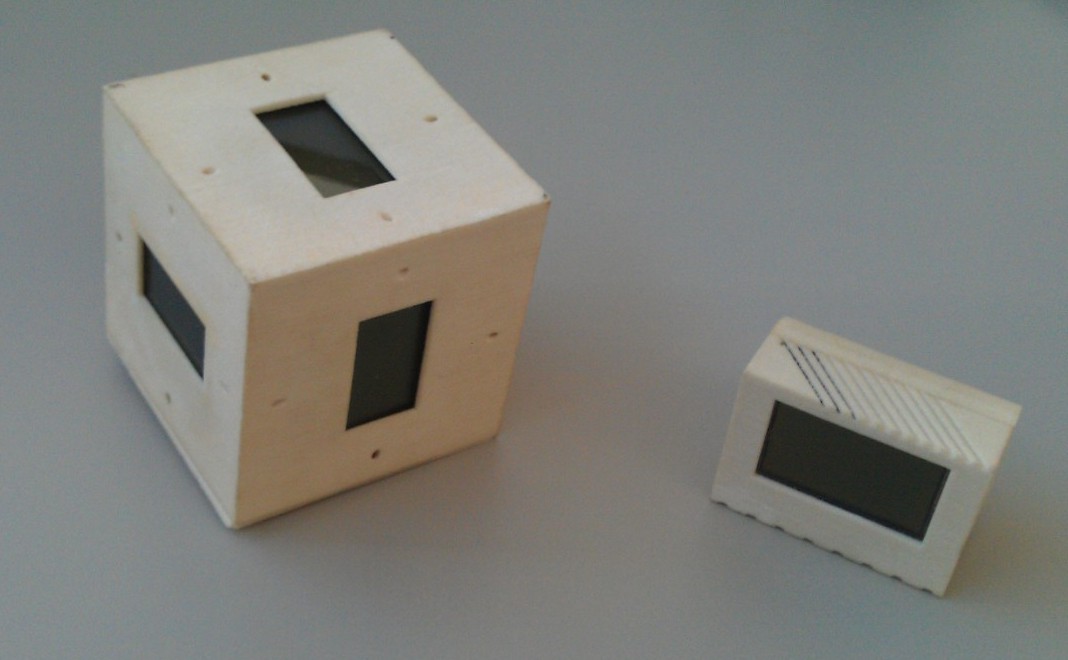}	
  \caption{The real counterparts to the virtual I/O cube and the presence
system vTUI depicted in Fig.~\ref{fig:virtual_tuis}. Due to the virtual hardware
drivers concept of the ROS middleware, the same software can be used for the
virtual and the real TUI, only the device drivers have to be replaced.}
  \label{fig:real_tuis}
\end{figure}

\section{Virtual TUI Examples}
In order to show the broad applicability of our toolkit, we have implemented simulations of several TUIs that have been
presented and published previously.

A collection of virtual TUIs is depicted in Fig.~\ref{fig:virtual_tuis}. It
shows (from left to right) two virtual \emph{Sifteo cubes}~\cite{Merrill2012},
a virtual I/O cube and a virtual TUI for a ubiquitous presence
system~\cite{kranz2006ubiquitouspresence}.
The meshes have been created with the free 3D
modeling software \emph{Blender}\footnote{\url{http://www.blender.org/}, last
accessed May\,12,
2014.}. In order to allow physical simulation, the
centers of gravity have been set to the objects' centers. Sizes and masses
correspond to the respective real objects.

All of these TUIs are equipped with virtual accelerometers and virtual
displays. The \emph{Sifteo cubes} further have virtual proximity sensors and
touch-sensitive screens (cf.\,Fig.~\ref{fig:real_sifteo}, right). The
communication system is based on the ROS
middleware's communication infrastructure that can also be used by real TUIs.

The left \emph{Sifteo cube} in Fig.~\ref{fig:virtual_tuis} is selected for
manipulation in Gazebo. The blocks and circles around the virtual object
represent the six degrees of freedom (DoF) that can be manipulated via the GUI.
Besides mouse and keyboard input, the manipulation can also be performed with
six-degrees-of-freedom (6DoF) technology such as 3Dconnexion's
SpaceNavigator\footnote{\url{
http://www.3dconnexion.com/products/spacenavigator.html}, last accessed
May\,18, 2014.}. This allows for a more realistic interaction with the 3D
scene.

Comparing to the fragile real objects (cf. Fig.~\ref{fig:real_tuis},
Fig.~\ref{fig:real_sifteo} left), new
interaction methods can also be explored with the virtual prototypes. Examples
are throwing the I/O cube like a dice or tossing the \emph{Sifteo cubes} like a
coin. In addition, it is possible to create virtual prototypes with sensors and
actuators that would be too expensive or are not yet available at a specific
size. This allows, for example, adding GPS sensors to tiny objects or using an
indoor positioning system with an accuracy that is not yet available today.

\begin{figure}[tb]
  \centering
  \includegraphics[width=0.48\textwidth]{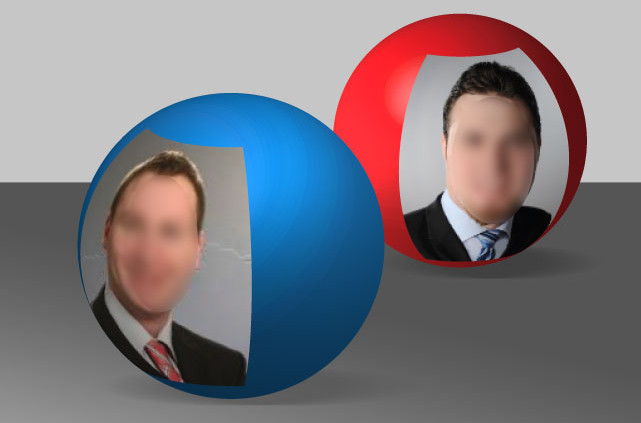}	
  \caption{Following Bishop's  Marble Answering Machine approach, we have
designed a virtual tangible messaging solution. In the simulation, we can
display images on a ball grows or shrinks depending on the amount of
messages and the time of last contact. For blind review, the faces have been blurred for reasons of anonymity.}
  \label{fig:virtual_marble}
\end{figure}

Following Bishop's Marble Answering Machine idea, we have designed a system
that allows managing personal messaging. Soft balls can display
images of contacts on their surfaces. Depending on the amount of exchanged
messages and the time elapsed since the last contact, the ball can grow or
shrink. When the ball is pressed, squeezed or bounced, the conversation pops up
on a tablet PC. The system is currently a prototype. The messaging applications
runs on a real tablet PC. The balls are simulated with our toolkit. The
communication is performed via the ROS middleware. The messaging application on
the Android-based tablet PC is using
\emph{rosjava}\footnote{\url{https://code.google.com/p/rosjava/}, last accessed
May 18, 2014.} for connecting with the balls. A virtual representation of two
balls is depicted in Fig.~\ref{fig:virtual_marble}. Creating a hardware
prototype of such a system would have been more costly and also have consumed more time and effort.

\section{Challenges Towards the Simulation\ \\of Tangible Aspects}

For shifting the whole prototyping phase of a TUI into simulation, users need
to be able to evaluate interactive as well as tangible aspects. However, with
the currently available input/output systems, the simulation is -- for the most
part -- limited to interactive aspects. This is mainly due to the lack of haptic
feedback. Many tangible interactions~\cite{Moller2012MobiMed}, such as squeezing a ball or feeling the
structure of a surface, cannot be experienced in the 3D environment. This often
restricts the evaluation of ``look and feel'' to the ``look'' part. Although our
experiences have shown that advanced 3D simulation users can also get a good
impression of the ``feel'' component over time, the assumed ``feeling'' from
advanced users cannot replace the user evaluations of TUIs.

The evaluation of tangible aspects could be enabled by introducing haptic
feedback that allows ``feeling'' the virtual objects and their surfaces. This
can be realized, for example, through haptic output devices such as wristbands
with vibration motors, fingertip tactile displays~\cite{WearableTactile2012} or
joint input/output devices such as a \emph{PHANTOM
Desktop}\footnote{\url{http://www.sensable.com/haptic-phantom-desktop.htm},
last accessed May\,17, 2014.}~(cf.~Wang et al.~\cite{Wang2012}).

For exploring 3D virtual scenes and manipulating virtual objects, the
mouse/keyboard combination is not optimal. By using a three-button mouse with
scroll wheel, the user can navigate in the 3D scene by performing the
translation and rotation movements one after the other. However, when it comes
to interaction, force can only be applied directly in the 2D plane the user has previously navigated
to, since the mouse is only a 2D input device. With the help of Gazebo's
6 DoF force targets (cf.\,Fig.~\ref{fig:virtual_tuis}, left object), this could
be overcome. However, applying force via predetermined targets does not
correspond to natural interaction. By using a 6 DoF device, the navigation can
be enormously simplified since translations and rotations can be done
simultaneously in a more natural way. However, applying force on an object is
still inconvenient since the target object has to be selected first, before the
user can give an impulsive to the object via the 6 DoF controller.

Exemplarily, we have examined the steps for spinning a top in the virtual
environment. In our example, we have used a 3Dconnexion SpaceNavigator together
with a standard mouse. The first step is navigating through the 3D environment
with the 6 DoF device until we reach the object and can see it from the
desired view point. For selecting the object which should be manipulated, we
need the mouse to activated the force control by clicking on the spinning top's
handle. In the third step, we can lift the spinning top by lifting the 6 DoF
device's control element. The last step is giving a rotational impulse via the
control element followed by an abrupt release of the device.

In order to simplify the exploration of the scene and the manipulation of
objects, for example, 3D navigation TUIs~\cite{WuTangibleNavigation2011} or
gesture interfaces~\cite{Liu2012} could be coupled to the simulation
environment. Combined with tactile displays, this would allow assessing the
tangible as well as the interactive aspect in a realistic way.

\section{Conclusion}

The proposed simulation approach based on the ROS middleware has several
advantages compared to classical prototyping approaches. For most developers,
the time savings will be the most important one. The possibility to
simulate tangible user interfaces with new and not yet realizable technologies
is another benefit. The effort in terms of costs and time to explore design
alternatives is significantly reduced. The interaction between real and
simulated devices allows extending available systems with novel devices.
Repeatable and easily modifiable test scenarios enable objective comparability
of different systems. Time and resources can also be saved for multi-device
scenarios, since an object can simply be spawned multiple times in the virtual
environment.

We extended the ROS middleware by several components that provide functions
necessary for TUIs. For example, we have developed components for display
outputs and various sensors, such as a touch sensor. By implementing
selected existing, previously published research prototypes of TUIs and a commercial platform, we confirmed the feasibility and function of the proposed approach.

Due to the limitations of the currently available off-the-shelf input/output
devices for 3D exploration, the proposed solution is not yet intended to fully
replace a physical prototype, but to minimize the time-consuming and costly
iterations for creating a working physical prototype. Future work includes
finding better suitable interfaces for exploring the 3D scene and manipulating
the virtual objects. In order to evaluate the benefit of our proposed solution,
we are currently setting up a exploratory study. The third-party participants
will be split up in two groups. One group will use a real hardware prototype-based
approach; the other group will perform prototyping with the virtual simulation.

\balance

\bibliographystyle{acm-sigchi}

\end{document}